\documentclass[pre,superscriptaddress,twocolumn,amsmath,amssymb,floatfix]{revtex4-1}

\usepackage{graphicx}

\begin{document}

\title{Measuring burstiness for finite event sequences} 
\author{Eun-Kyeong Kim}
\affiliation{GeoVISTA Center, Department of Geography, Pennsylvania State University, PA 16802, USA}
\author{Hang-Hyun Jo}
\email{johanghyun@postech.ac.kr}
\affiliation{BK21plus Physics Division and Department of Physics, Pohang University of Science and Technology, Pohang 37673, Republic of Korea}
\affiliation{Department of Computer Science, Aalto University School of Science, P.O. Box 15500, Espoo, Finland}
\date{\today}

\begin{abstract}
Characterizing inhomogeneous temporal patterns in natural and social phenomena is important to understand underlying mechanisms behind such complex systems, hence even to predict and control them. Temporal inhomogeneities in event sequences have been described in terms of bursts that are rapidly occurring events in short time periods alternating with long inactive periods. The bursts can be quantified by a simple measure, called burstiness parameter, which was introduced by Goh and Barab\'asi [EPL \textbf{81}, 48002 (2008)]. The burstiness parameter has been widely used due to its simplicity, which however turns out to be strongly affected by the finite number of events in the time series. As the finite-size effects on burstiness parameter have been largely ignored, we analytically investigate the finite-size effects of the burstiness parameter. Then we suggest an alternative definition of burstiness that is free from finite-size effects and yet simple. Using our alternative burstiness measure, one can distinguish the finite-size effects from the intrinsic bursty properties in the time series. We also demonstrate the advantages of our burstiness measure by analyzing empirical datasets.
\end{abstract}



\maketitle

\section{Introduction}

Characterizing inhomogeneous temporal patterns in natural and social phenomena~\cite{Utsu1995Centenary, Wheatland1998WaitingTime, deArcangelis2006Universality, Kemuriyama2010Powerlaw, Barabasi2005Origin, Vazquez2006Modeling} is important to understand underlying mechanisms behind such complex systems, hence even to predict and control them. Recently, temporal inhomogeneities in event sequences have been described by bursts that are rapidly occurring events in short time periods alternating with long inactive periods~\cite{Barabasi2005Origin}. Bursty activity patterns of elements in a complex system turn out to strongly affect the collective dynamics emerging from interaction between elements, such as epidemic spreading or diffusion~\cite{Vazquez2007Impact, Karsai2011Small, Miritello2011Dynamical, Iribarren2011Branching, Rocha2011Simulated, Starnini2012Random, VanMieghem2013NonMarkovian, Jo2014Analytically, Delvenne2015Diffusion}. Despite such a crucial role of bursts in complex systems, a quantification of burstiness remains largely unexplored.

Bursts for a given event sequence have often been characterized in terms of a heavy-tailed interevent time distribution $P(\tau)$, where the interevent time $\tau$ is defined as the time interval between two consecutive events. The interevent time distribution could be even reduced to a single number $B$, called burstiness parameter~\cite{Goh2008Burstiness}:
\begin{equation}
    \label{eq:originalB}
    B= \tfrac{\sigma-\mu}{\sigma+\mu}=\tfrac{r-1}{r+1},
\end{equation}
where $\sigma$ and $\mu$ denote the standard deviation and the mean of interevent times, respectively, and $r=\frac{\sigma}{\mu}$ is the coefficient of variation. $B$ has the value of $-1$ for regular time series as $\sigma=0$, and $0$ for Poissonian or random time series as $\sigma=\mu$. Finally, the value of $B$ approaches $1$ for extremely bursty time series as $\sigma\to\infty$ for finite $\mu$. This burstiness parameter has been widely used mainly due to its simplicity, e.g., to analyze earthquake records, heartbeats of human subjects, and mobile phone communication patterns as well as to test models for bursty dynamics~\cite{Goh2008Burstiness, Jo2012Circadian, Yasseri2012Dynamics, Kim2014Index, Zhao2015Empirical, Jo2015Correlated, Gandica2016Origin}. $r$ has been also used for investigating spatiotemporal organization of aftershocks in seismology~\cite{Bottiglieri2009Identification}. In addition, higher order correlations between interevent times and long-range memory effects have been studied in terms of memory coefficient~\cite{Goh2008Burstiness}, autocorrelation function or power spectrum~\cite{Jo2012Circadian, Karsai2012Universal}, and bursty train distribution~\cite{Karsai2012Universal}. 

We note that the behavior of burstiness parameter in Eq.~(\ref{eq:originalB}) may not be robust with respect to finite-size effects. The behavior of $B$ approaching $1$ is expected only when the number of events in the time series is sufficiently large or infinite. In other words, the maximum value of $B$ is strongly limited by the number of events. However, the numbers of events in datasets are finite for real systems, and they are often too small to draw statistically significant conclusions from data analysis. Further, the number of events or activity level per individual is typically highly skewed, e.g., in human dynamics~\cite{Radicchi2009Human, Jo2012Circadian}, implying that the majority of population has relatively small number of events. Those individuals with low activity have been arbitrarily ignored, or aggregated to form a group of low activity. It could also be due to the absence of reliable measures of burstiness for finite event sequences. Thus, the questions arise: How can one properly compare one bursty time series with another that shows similar bursty pattern but has the different number of events? In other words, how can one isolate finite-size effects from the intrinsic temporal features in time series?

In order to study finite-size effects on burstiness parameter, we devise an analytically tractable model to calculate the coefficient of variation of interevent times for finite event sequences in Section~\ref{sec:model}. Our model has two relevant factors, i.e., a bursty period and a lower bound of interevent time. By tuning these two control parameters, we can obtain the analytic values of $B$ for three reference cases; regular, random, and extremely bursty time series. Then we investigate the strong finite-size effects on $B$ for reference cases, enabling us to suggest an alternative definition of burstiness parameter that is free from finite-size effects and yet simple. We also demonstrate the advantages of our alternative burstiness parameter by analyzing empirical datasets. Finally, we conclude our work in Section~\ref{sec:conclusion}.

\section{Model} \label{sec:model}

\subsection{Uniform case}\label{subsec:uniform}

We first consider an event sequence with $n$ events, each taking place uniformly at random in the time interval $[0,d)$. The events are ordered by their timings, and the timing of the $i$th event is denoted by $t_i$ for $i=1,\cdots,n$. We consider the case of periodic boundary condition in time for simplicity, while the case of open boundary condition is discussed in Appendix~\ref{sect:OBC}. Interevent times are defined as
\begin{equation}
    \label{eq:tau_d}
    \tau_{i,d}\equiv \left\{\begin{tabular}{ll}
    $d-t_n+t_1$ & if $i=1$\\
    $t_i-t_{i-1}$ & if $i\neq 1$.
  \end{tabular}\right.
\end{equation}
By the order statistics~\cite{David2003Order, Kivela2012Multiscale}, interevent time distributions are written as follows:
\begin{equation}
    \label{eq:Ptau_id}
    P(\tau_{i,d})=\left\{\begin{tabular}{ll}
      $\frac{(\tau_{1,d}/d)(1-\tau_{1,d}/d)^{n-2}}{B(2,n-1)d}$ & if $i=1$\\
      $\frac{(1-\tau_{i,d}/d)^{n-1}}{B(1,n)d}$ & if $i\neq 1$,\\
      \end{tabular}\right.
\end{equation}
where $B(n,m)$ denotes the beta function, 
\begin{equation}
B(n,m)=\int_0^1 z^{n-1}(1-z)^{m-1}dz=\tfrac{(n-1)!(m-1)!}{(n+m-1)!}.
\end{equation}
Expectation values of $\tau_{i,d}$ and $\tau^2_{i,d}$ are obtained as
\begin{equation}
    \langle\tau_{i,d}\rangle=\left\{\begin{tabular}{ll}
    $\frac{2d}{n+1}$ & if $i=1$\\
    $\frac{d}{n+1}$ & if $i\neq 1$,\\
  \end{tabular}\right.
\end{equation}
and
\begin{equation}
    \langle\tau^2_{i,d}\rangle=\left\{\begin{tabular}{ll}
    $\frac{6d^2}{(n+1)(n+2)}$ & if $i=1$\\
    $\frac{2d^2}{(n+1)(n+2)}$ & if $i\neq 1$.\\
  \end{tabular}\right.
\end{equation}
We have assumed that $\tau_{i,d}$s are independent of each other, although they are not independent but to satisfy the condition $\sum_{i=1}^n \tau_{i,d}=d$. Instead we find on average that 
\begin{equation}
    \sum_{i=1}^n\langle \tau_{i,d}\rangle= \langle \tau_{1,d}\rangle +(n-1) \langle \tau_{i\neq 1,d}\rangle=d.
\end{equation}
This issue will be discussed later in the next Subsection.

\subsection{Localized model}\label{subsec:local}

\begin{figure}[!t]
    \includegraphics[width=\columnwidth]{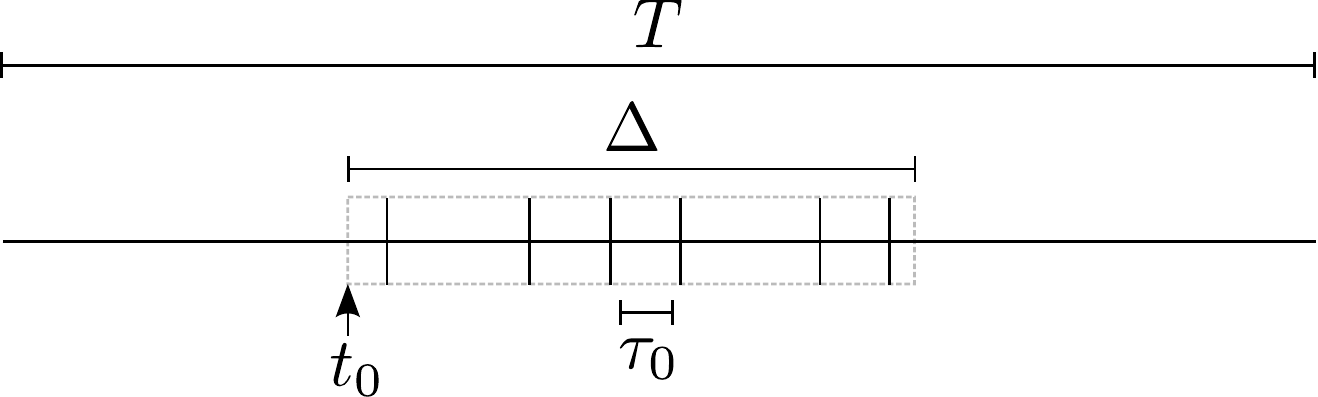}
    \caption{Schematic diagram of the localized model: $n$ events are localized in the period $\Delta$ beginning at $t_0$ in $[0,T)$, and they are separated from each other at least by $\tau_0$.}
\label{fig:model}
\end{figure}

We now consider the general case that all events are localized in the interval $[t_0,t_0+\Delta)$ with $t_0\geq 0$ and $t_0+\Delta<T$, indicating that events do not take place in the intervals $[0,t_0)$ and $[t_0+\Delta,T)$, as depicted in Fig.~\ref{fig:model}. A similar model has been studied in a different context~\cite{Perotti2014Temporal}. The localization parameter $\Delta$ is introduced to simulate the bursty limit for $\Delta\ll T$. Since we use periodic boundary condition, $t_0$ can be ignored. In addition, the lower bound of interevent time, $\tau_0$, is introduced, implying that events must be separated from each other at least by $\tau_0$. Accordingly, it is assumed that
\begin{equation}
    \label{eq:condition_DeltaTau}
(n-1)\tau_0\leq \Delta\leq T-\tau_0,
\end{equation}
leading to $\tau_0\leq \tfrac{T}{n}$. If $\tau_0=\tfrac{T}{n}$, one gets the regular time series.

Then, we use definitions in Eq.~(\ref{eq:tau_d}) with $d$ being replaced by $\delta\equiv\Delta-(n-1)\tau_0$ to define interevent times as
\begin{equation}
  \tau_i\equiv \left\{\begin{tabular}{ll}
          $\tau_{1,\delta}+T-\Delta$ & if $i=1$\\
          $\tau_{i,\delta}+\tau_0$ & if $i\neq 1$.\\
  \end{tabular}\right.
\end{equation}
Using Eq.~(\ref{eq:Ptau_id}), we get
\begin{equation}
  \langle\tau_i\rangle=\left\{\begin{tabular}{ll}
          $T-\frac{n-1}{n+1}(\Delta+2\tau_0)$ & if $i=1$\\
    $\frac{\Delta+2\tau_0}{n+1}$ & if $i\neq 1$,\\
  \end{tabular}\right.
\end{equation}
and
\begin{equation}
  \langle\tau_i^2 \rangle=
  \left\{\begin{tabular}{ll}
    $\tfrac{6\delta^2}{(n+1)(n+2)}+\tfrac{4(T-\Delta)\delta}{n+1}+(T-\Delta)^2$ & if $i=1$\\
    $\tfrac{2\delta^2}{(n+1)(n+2)}+\tfrac{2\tau_0\delta}{n+1}+\tau_0^2$ & if $i\neq 1$.\\
  \end{tabular}\right.
\end{equation}
Then we calculate the mean $\mu_n$ and the variance $\sigma^2_n$ of interevent times to get the coefficient of variation $r_n=\frac{\sigma_n}{\mu_n}$:
\begin{eqnarray}
    \mu_n &=& \tfrac{1}{n}[\langle\tau_1\rangle+(n-1)\langle \tau_{i\neq 1}\rangle],\\
    \sigma^2_n &=& \tfrac{1}{n}[\langle\tau_1^2\rangle+(n-1)\langle \tau_{i\neq 1}^2\rangle]-\mu_n^2,\\
    \label{eq:r_n}
    r_n(x,y) &=& \sqrt{\tfrac{(n-1)[1+n(1-x)^2+n(n+1)y^2-2n(2-x)y]}{n+1}}.
\end{eqnarray}
Here we have defined
\begin{equation}
    x\equiv \tfrac{\Delta}{T},\ y\equiv \tfrac{\tau_0}{T},
\end{equation}
satisfying the conditions that 
\begin{equation}
    \label{eq:condition_xy}
    (n-1)y\leq x\leq 1-y,\ y\leq \tfrac{1}{n}.
\end{equation}
It is straightforward to show that $r_n(x,y)$ is a non-increasing function of $x$ and $y$, respectively.

\begin{figure}[!t]
    \includegraphics[width=\columnwidth]{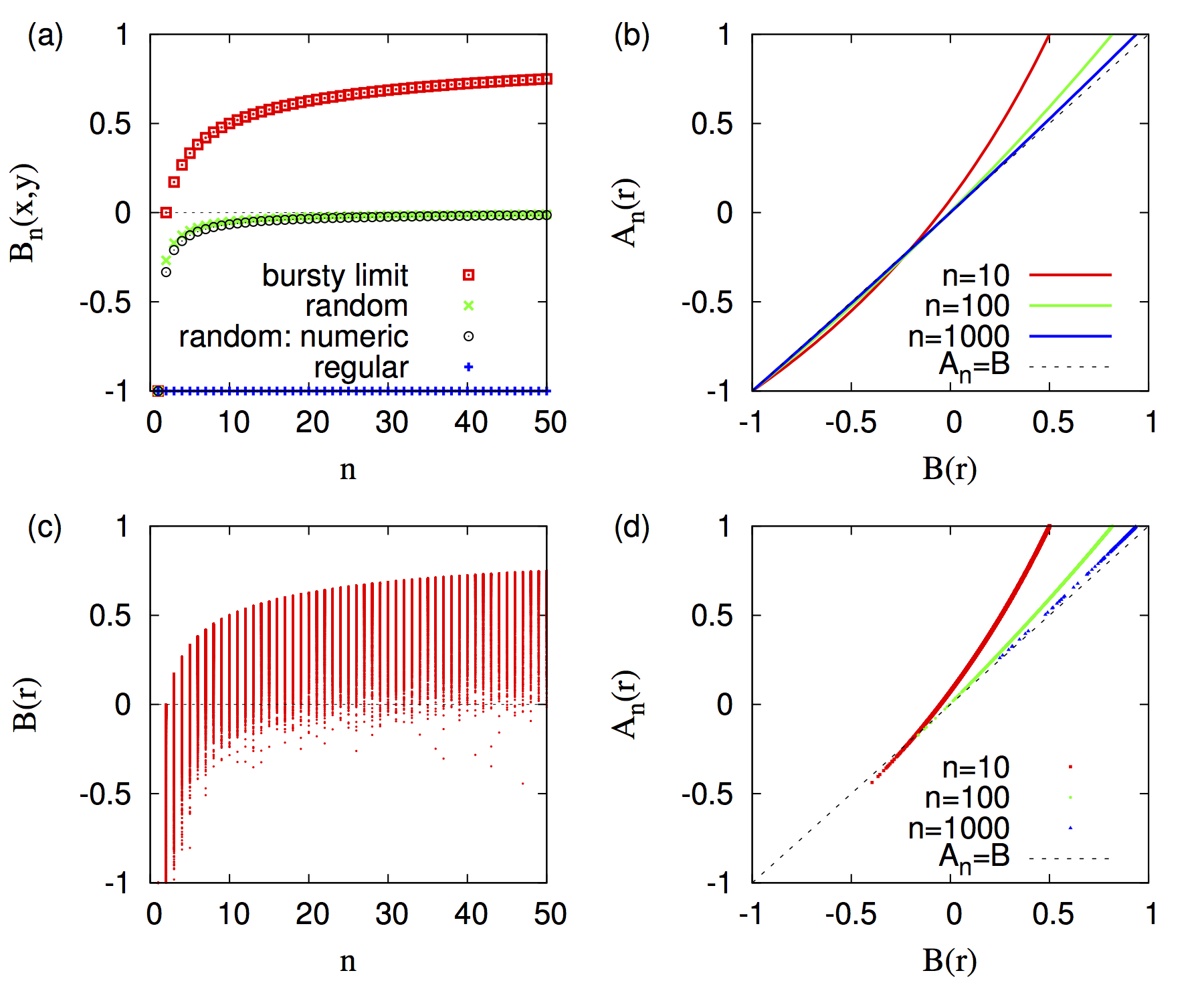}
    \caption{(Color online) (a) Analytic results of $B_n(x,y)$ for three reference cases: Eq.~(\ref{eq:regular}) for regular time series, Eq.~(\ref{eq:Poisson}) for random time series, and Eq.~(\ref{eq:bursty}) for the bursty limit. Numerical results for the random case are plotted for comparison to the analytic results. (b) Comparison of the novel burstiness measure $A_n(r)$ in Eq.~(\ref{eq:Anr}) to the original burstiness parameter $B(r)$ in Eq.~(\ref{eq:originalB}) for several values of $n$. (c) Scatter plot of $B(r)$ for individual Twitter users. (d) The same as (b) but using Twitter dataset.}
\label{fig:ABn_pbc}
\end{figure}

In order to study the strong finite-size effects in event sequences, we define the burstiness parameter using $x$ and $y$ as follows:
\begin{equation}
    B_n(x,y)\equiv\tfrac{r_n(x,y)-1}{r_n(x,y)+1}.
\end{equation}
We discuss three reference cases. Firstly, the regular time series means that all interevent times are the same as $\mu_n$, implying that $x=1-\frac{1}{n}$ and $y=\frac{1}{n}$. Since $r_n=0$ independent of $n$, we get
\begin{equation}
    \label{eq:regular}
    B_n(1-\tfrac{1}{n},\tfrac{1}{n})=-1.
\end{equation}
Secondly, the Poissonian or random time series corresponds to the case with $\Delta=T$ and $\tau_0=0$, i.e., $x=1$ and $y=0$, leading to $r_n=\sqrt{\tfrac{n-1}{n+1}}$. We get
\begin{equation}
    \label{eq:Poisson}
    B_n(1,0)=\tfrac{\sqrt{n-1}-\sqrt{n+1}}{\sqrt{n-1}+\sqrt{n+1}}.
\end{equation}
Note that $B_1(1,0)=-1$ and that $B_n(1,0)$ is always negative but approaches $0$ as $n$ increases, i.e., $B_n(1,0) \approx -\tfrac{1}{2n}$ for large $n$, as shown in Fig.~\ref{fig:ABn_pbc}(a). Since this result is based on the assumption of independence of $\tau_i$s, we test our result by comparing it to numerical values of burstiness parameter. For this, we generate $10^5$ event sequences for each $n$ to obtain the burstiness parameter as depicted in Fig.~\ref{fig:ABn_pbc}(a). We find that the deviation of our analytic results from the simulations is negligible.

Finally, the extremely bursty time series corresponds to the case that all events occur asymptotically at the same time, i.e., $x=y=0$, leading to $r_n=\sqrt{n-1}$. Thus one gets 
\begin{equation}
    \label{eq:bursty}
    B_n(0,0)=\tfrac{\sqrt{n-1}-1}{\sqrt{n-1}+1}.
\end{equation}
Note that $B_1(0,0)=-1$ and $B_2(0,0)=0$. $B_n(0,0)$ becomes positive for $n\geq 3$, and then approaches $1$ as $n$ increases, i.e., $B_n(0,0)\approx 1-\tfrac{2}{\sqrt{n}}$, as shown in Fig.~\ref{fig:ABn_pbc}(a). The finite-size effect turns out to be the strongest for the bursty case. It is because $B=1$ is realized only for infinitely many interevent times. The strong dependence of $B_n$ on the number of events $n$ could introduce a serious finite-size effects in particular for the bursty case, i.e., for most empirical datasets.

We remark that for our localized model, only one burst with period $\Delta$ is assumed to exist. This assumption is sufficient to simulate above reference cases. It is because even when considering multiple bursts in the event sequence, above reference cases will be the same as those in our localized model. In any case, the extension of our model to incorporate multiple bursts could be important, but left for future works.

\subsection{Novel definition of burstiness measure}\label{subsec:newB}

In order to fix the finite-size effects in the original burstiness parameter, we suggest a novel definition of the burstiness measure, denoted by $A_n(r)$~\footnote{As our novel definition of burstiness $A_n(r)$ takes the sample size $n$ into account, it is not a parameter but it could be interpreted as an unbiased estimator for $B(r)$ in the statistical sense. We however call $A_n(r)$ a measure mainly because it has been proposed as the descriptive statistic rather than as an optimal unbiased estimator of $B(r)$. Note that finding an unbiased estimator of coefficient of variation $r$ is not trivial at all~\cite{Sokal1980Significance, Breunig2001Almost, Forkman2009Estimator, Albrecher2010Asymptotics, Banik2012Testing, Jayakumar2015Exact, Kivela2015Estimating}, hence beyond the scope of our current work. Thus, statistical evaluation of the error of our novel definition $A_n(r)$ is left for a future work.}. This burstiness meausre is assumed to be a function of $r=\tfrac{\sigma_n}{\mu_n}$. Then $A_n(r)$ must satisfy the following conditions:
\begin{eqnarray}
    A_n(0) &=& -1,\nonumber \\
    A_n\left(\sqrt{\tfrac{n-1}{n+1}}\right) &=& 0,\label{eq:conditions_An}\\
    A_n\left(\sqrt{n-1}\right) &=& 1,\nonumber
\end{eqnarray}
which correspond to the cases of regular, random, and extremely bursty time series, respectively. Since $B(r)$ was originally defined as $\tfrac{r-1}{r+1}$, we similarly assume that $A_n(r)=\tfrac{a_nr-b_n}{r+c_n}$ with coefficients $a_n$, $b_n$, and $c_n$. Using a general formula of Eq.~(\ref{eq:generalA}) in Appendix~\ref{sect:generalA}, we get 
\begin{equation}
    \label{eq:Anr}
    A_n(r)=\tfrac{\sqrt{n+1}r-\sqrt{n-1}}{(\sqrt{n+1}-2)r+\sqrt{n-1}}
\end{equation}
for $0\leq r\leq \sqrt{n-1}$. Our novel burstiness measure $A_n$ has no longer an upper bound due to the finite $n$, as depicted in Fig.~\ref{fig:ABn_pbc}(b), where the curves for different $n$s are described by
\begin{equation}
    A_n(r)=\tfrac{ \sqrt{n+1}-\sqrt{n-1}+(\sqrt{n+1}+\sqrt{n-1})B(r)} {\sqrt{n+1}+\sqrt{n-1}-2+(\sqrt{n+1}-\sqrt{n-1}-2)B(r)}.
\end{equation}
Then let us consider two event sequences with the same value of $r$ but with different $n$s. The original burstiness parameter $B$ has the same value, while $A_n$ gives different values. Thus, $A_n$ can reveal the difference in bursty patterns without finite-size effects.

For practical applications, we show how $A_n$ can be used to quantify the burstiness in the empirical dataset without finite-size effects. For a given event sequence of $n$ events, one can calculate the coefficient of variation of interevent times, denoted by $\tilde r$, to get the value of $A_n(\tilde r)$ for the given event sequence. For the demonstration, we analyze a large-scale Twitter dataset collected for a year from October 1, 2012 to September 30, 2013 throughout the United States of America~\footnote{The dataset was originally collected only for tweets with geographical information, while we exploit only the temporal information of tweets in our work.}. After cleaning, the dataset contains approximately $698$ million tweets posted by about $5.5$ million user accounts. As shown in Fig.~\ref{fig:ABn_pbc}(c), the original burstiness parameter for individual Twitter users in the dataset clearly has the upper bound for small values of $n$. Such upper bound is removed or corrected in $A_n$ obtained from the same dataset in Fig.~\ref{fig:ABn_pbc}(d).

\begin{figure}[!t]
    \includegraphics[width=\columnwidth]{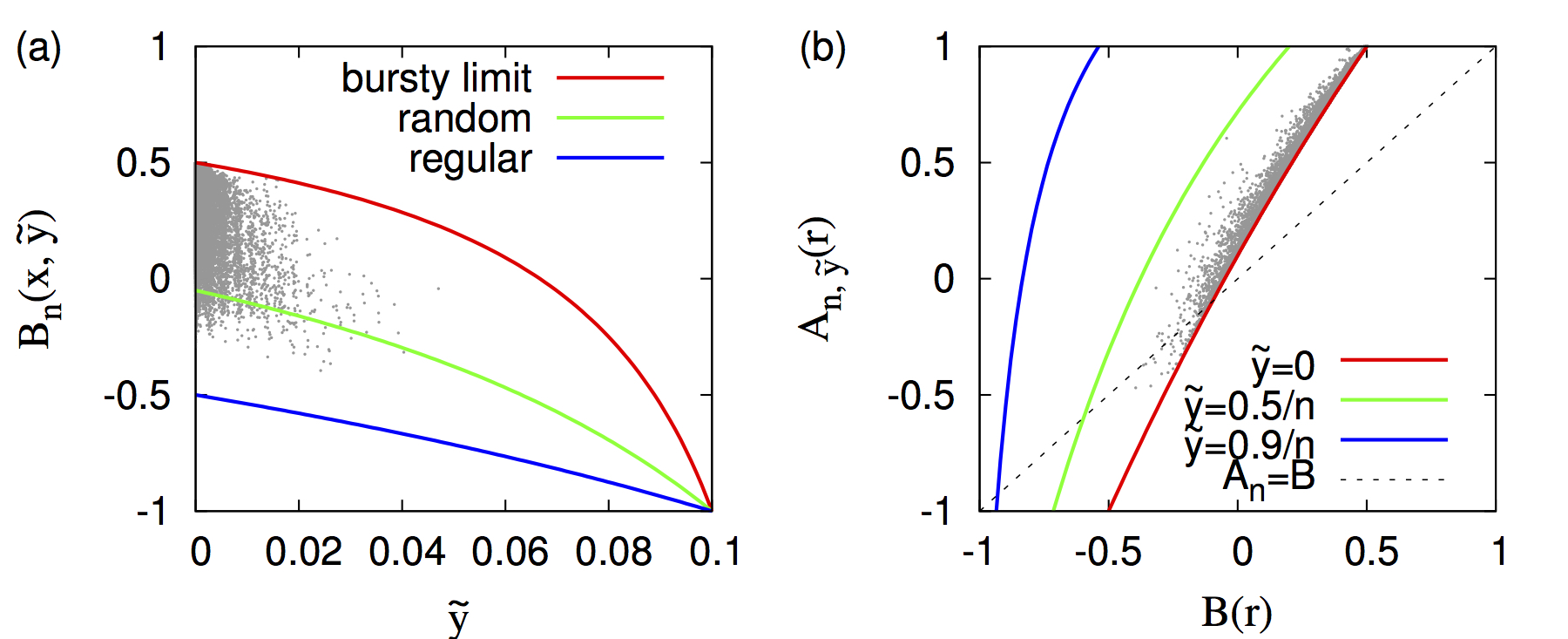}
    \caption{(Color online) (a) Analytic results of $B_n(x,\tilde y)$ for bursty limit and random cases, and that of $B_n(r^*_n(\tilde y))$ for regular case, when $n=10$. (b) Comparison of the novel burstiness measure $A_{n,\tilde y}(r)$ in Eq.~(\ref{eq:Anyr}) to the original burstiness parameter $B(r)$ in Eq.~(\ref{eq:originalB}) for several values of $\tilde y$ for $n=10$. Each gray dot in both panels corresponds to each individual Twitter user with $n=10$.}
\label{fig:ABny_pbc_n10}
\end{figure}

\subsection{Effect due to minimum interevent times}\label{subsec:mintau}

In addition to the number of events in a given event sequence, one can also exploit more information from the sequence, such as the minimum interevent time, $\tau_{\rm min}=\min \{\tau_i\}$. The role of the minimum interevent time has been discussed in various contexts. For example, it can be related to the refractory period of neurons, which may limit the response time of neuronal systems to external stimuli. The minimum interevent time has been found to play a crucial role in spreading dynamics in complex systems~\cite{Jo2014Analytically}. Since different systems or different individuals in the same system may have different values of minimum interevent time, its effect must be carefully investigated, particularly for comparing the temporal properties of different systems or different individuals in the same system.

We consider three reference cases for given $\tau_{\rm min}$ or $\tilde y\equiv \tfrac{\tau_{\rm min}}{T}$. The bursty limit case is achieved by maximizing $r_n(x,\tilde y)$ with $x=(n-1)\tilde y$, see Eq.~(\ref{eq:condition_xy}). We get the random case by setting $x=1-\tilde y$ for $r_n(x,\tilde y)$. Finally, as for the regular case we consider a specific time series that one interevent time is $\tau_{\rm min}$, while all other $n-1$ interevent times are the same as $\frac{T-\tau_{\rm min}}{n-1}$ for $n\geq 2$. We calculate the coefficient of variation of interevent times as $r^*_n(\tilde y)=\frac{1-n\tilde y}{\sqrt{n-1}}$~\footnote{Other regular cases can be considered. For example, $k$ interevent times are the same as $\tau_{\rm min}$, while other $n-k$ interevent times are the same as $\frac{T-\tau_{\rm min}}{n-k}$ for $n>k$. Then we get the coefficient of variation as $\sqrt{\frac{k}{n-k}}(1-n\tilde y)$, indicating that the minimal coefficient of variation is obtained when $k=1$.}. Note that $\tau_0$ in the localized model is the possible lower bound of interevent time, while $\tau_{\rm min}$ is one of interevent times in the given event sequence, hence it can be larger than $\tau_0$. In sum, we get
\begin{eqnarray}
    r^*_n(\tilde y) &=& \tfrac{1-n\tilde y}{\sqrt{n-1}},\\
    r_n(1-\tilde y,\tilde y) &=& \sqrt{\tfrac{n-1}{n+1}}(1-n\tilde y),\\
    r_n((n-1)\tilde y,\tilde y) &=& \sqrt{n-1}(1-n\tilde y)
\end{eqnarray}
for the cases of regular, random, and extremely bursty time series, respectively. Since the specific time series for the regular case does not fit within the localized model, we need to impose the condition that $r^*_n(\tilde y)< r_n(1-\tilde y,\tilde y)$, i.e., $n>3$. One can calculate $B(r)=\frac{r-1}{r+1}$ with above results of $r$s for reference cases to find the strong effects due to the finite minimum interevent time and the finite size of events. For example, the analytic curves for $n=10$ are depicted in Fig.~\ref{fig:ABny_pbc_n10}(a), where we also plot the empirical values of $B$ for individual Twitter users with $n=10$ according to their own $\tilde y$.

Then, one can find the functional form of the burstiness measure $A_{n,\tilde y}(r)=\tfrac{a_{n,\tilde y}r-b_{n,\tilde y}}{r+c_{n,\tilde y}}$ with coefficients $a_{n,\tilde y}$, $b_{n,\tilde y}$, and $c_{n,\tilde y}$, satisfying
\begin{eqnarray}
    A_{n,\tilde y}[r^*_n(\tilde y)]&=&-1,\nonumber\\
    A_{n,\tilde y}[r_n(1-\tilde y,\tilde y)]&=&0,\label{eq:conditions_Any}\\
    A_{n,\tilde y}[r_n((n-1)\tilde y,\tilde y)]&=&1.\nonumber
\end{eqnarray}
Using a general formula of Eq.~(\ref{eq:generalA}) in Appendix~\ref{sect:generalA}, we obtain the complete form as following:
\begin{equation}
    \label{eq:Anyr}
    A_{n,\tilde y}(r)=\tfrac{ (n-2)[\sqrt{n+1}r-\sqrt{n-1}(1-n\tilde y)] } { [n\sqrt{n+1}-2(n-1)] r + \sqrt{n-1}(n-2\sqrt{n+1})(1-n\tilde y) }
\end{equation}
for $\frac{1-n\tilde y}{\sqrt{n-1}}\leq r\leq \sqrt{n-1}(1-n\tilde y)$. In Fig.~\ref{fig:ABny_pbc_n10}(b), we compare $A_{n,\tilde y}(r)$ against $B(r)$ for the entire range of $r$, with empirical results for Twitter users with $n=10$. We remark that for given event sequences, one can exploit even more information from those sequences, depending on which factors are to be controlled.

\section{Conclusions}\label{sec:conclusion}

Recently, temporal inhomogeneities in various natural and social event sequences have been described in terms of bursts, which are rapidly occurring events in short time periods separated by long inactive periods. Bursts have been mostly characterized by heavy-tailed interevent time distributions, or more simply by the burstiness parameter $B(r)=\frac{r-1}{r+1}$, where the coefficient of variation, $r$, denotes the ratio of standard deviation to mean of interevent times. $B$ has the value of $-1$ for regular time series as $r=0$, and it is $0$ for Poissonian or random time series as $r=1$. Finally, $B(r)$ approaches $1$ for extremely bursty time series as $r\to\infty$. Despite its successful applications, $B(r)$ turns out to be strongly affected by the finite size of event sequence, in particular when the event sequence is bursty. In order to get the analytic limits of $B(r)$ for the given $n$ in the reference cases, i.e., regular, random, and extremely bursty cases, we devise and study an analytically tractable model with $n$ events. Then we suggest a novel definition of burstiness measure that is free from finite-size effects and yet simple, denoted by $A_n(r)$: $A_n(r)$ has no upper bound due to the finite $n$. If two event sequences have the same $r$ but different $n$s, the original burstiness parameter cannot distinguish which event sequence is more bursty than the other, while our novel burstiness measure can do. Thus one can isolate the effect due to the finite size of event sequences. By analyzing a large-scale Twitter dataset, we show that $B(r)$ clearly has the upper bound due to the finite $n$ for individual Twitter users, and that $A_n(r)$ has no longer such upper bound.

We can exploit more information other than $n$ from the given event sequences, such as the minimum interevent time. The minimum interevent time is important to understand the intrinsic timescale of the system, e.g., the refractory period of neurons. For fixing the effects due to the finite minimum interevent time and the finite size of event sequence, we suggest another burstiness measure $A_{n,\tilde y}(r)$ with $\tilde y$ denoting the ratio of minimum interevent time to the whole period. Using the $A_{n,\tilde y}(r)$, we can separate the intrinsic bursty dynamics from the effect due to the minimal timescale in the system. Even more information in the given event sequences can be exploited, which we leave for future works. 

We have considered only the burstiness parameter, while other quantities like memory coefficient~\cite{Goh2008Burstiness} and bursty train distribution~\cite{Karsai2012Universal} for higher order temporal correlations could be also investigated from the same perspective. In addition, our analytically tractable model can be extended to incorporate another kind of information called context, e.g., as studied in terms of contextual bursts~\cite{Jo2013Contextual}.

\begin{acknowledgments}
    We would like to thank Mikko Kivel\"a and Hyokun Yun for helpful discussions, and anonymous reviewers for comments on our manuscript. E.-K. K. acknowledges the access to geo-tagged Twitter data collected by the group of Prof. Marcel Salath\'e at the \'Ecole Polytechnique F\'ed\'erale de Lausanne, Switzerland, and advice and useful comments from Alan M. MacEachren at the Pennsylvania State University, USA. H.-H.~J. acknowledges financial support by Basic Science Research Program through the National Research Foundation of Korea (NRF) grant funded by the Ministry of Education (2015R1D1A1A01058958).
\end{acknowledgments}

\appendix

\section{Open boundary condition}\label{sect:OBC}

Here we obtain the analytic results in the case of open boundary condition as the periodic boundary condition may not apply to some empirical datasets. 

We first consider an event sequence with $n$ events, each taking place uniformly at random in the interval $[0,d)$. The events are ordered by their timings, and the timing of the $i$th event is denoted by $t_i$ for $i=1,\cdots,n$. Then, interevent times are defined as
\begin{equation}
    \label{eq:OBC_tau_d}
    \tau_{i,d}=\left\{\begin{tabular}{ll}
    $t_1$ & if $i=1$\\
    $t_i-t_{i-1}$ & if $i=2,\cdots,n$\\
    $d-t_n$ & if $i=n+1$.
  \end{tabular}\right.
\end{equation}
Here the time intervals from $t=0$ to $t=t_1$ and from $t=t_n$ to $t=d$ have been taken as ``interevent times", although there is no event at $t=0$ and $t=d$. The case when these time intervals are discarded is left for future works. By the order statistics~\cite{David2003Order}, interevent time distributions read as follows:
\begin{equation}
    \label{eq:OBC_Ptau_id}
    P(\tau_{i,d})= \tfrac{(1-\tau_{i,d}/d)^{n-1}}{B(1,n)d}
\end{equation}
for all $i$. Expectation values of $\tau_{i,d}$ and $\tau^2_{i,d}$ are obtained as
\begin{eqnarray}
    \langle\tau_{i,d}\rangle &=& \tfrac{d}{n+1},\\
    \langle\tau_{i,d}^2\rangle &=& \tfrac{2d^2}{(n+1)(n+2)}.
\end{eqnarray}
Here the interevent times satisfy $\sum_{i=1}^{n+1}\langle \tau_{i,d}\rangle = d$. Thus, one gets
\begin{eqnarray}
    \mu_n &=& \tfrac{d}{n+1},\\
    \sigma^2_n &=& \tfrac{nd^2}{(n+1)^2(n+2)}.
\end{eqnarray}

We then consider the general case that all events are localized in the interval $[t_0,t_0+\Delta)$ with $t_0\geq 0$ and $t_0+\Delta<T$, as depicted in Fig.~\ref{fig:model}. In addition to the condition in Eq.~(\ref{eq:condition_DeltaTau}), we have one more condition for $t_0$ that
\begin{equation}
    \tau_0\leq t_0\leq T-\Delta-\tau_0.
\end{equation}
We use definitions in Eq.~(\ref{eq:OBC_tau_d}) with $d$ being replaced by $\delta\equiv\Delta-(n-1)\tau_0$ to define interevent times as
\begin{equation}
  \tau_i\equiv \left\{\begin{tabular}{ll}
          $\tau_{1,\delta}+t_0$ & if $i=1$\\
          $\tau_{i,\delta}+\tau_0$ & if $i=2,\cdots,n$\\
          $\tau_{n+1,\delta}+T-\Delta-t_0$ & if $i=n+1$.\\
  \end{tabular}\right.
\end{equation}
Using Eq.~(\ref{eq:OBC_Ptau_id}), one gets
\begin{equation}
    \langle \tau_i\rangle = \left\{\begin{tabular}{ll}
          $\tfrac{\delta}{n+1}+t_0$ & if $i=1$\\
          $\tfrac{\delta}{n+1}+\tau_0$ & if $i=2,\cdots,n$\\
          $\tfrac{\delta}{n+1}+T-\Delta-t_0$ & if $i=n+1$.\\
  \end{tabular}\right.
\end{equation}
and
\begin{eqnarray}
  &&\langle\tau_i^2 \rangle = \nonumber\\
  &&\left\{\begin{tabular}{ll}
    $\tfrac{2\delta^2}{(n+1)(n+2)}+\tfrac{2t_0\delta}{n+1}+t_0^2$ & if $i=1$\\
    $\tfrac{2\delta^2}{(n+1)(n+2)}+\tfrac{2\tau_0\delta}{n+1}+\tau_0^2$ & if $i=2,\cdots,n$\\
    $\tfrac{2\delta^2}{(n+1)(n+2)}+\tfrac{2(T-\Delta-t_0)\delta}{n+1}+(T-\Delta-t_0)^2$ & if $i=n+1$.
  \end{tabular}\right.\nonumber\\
\end{eqnarray}
\begin{widetext}
The mean and the variance of interevent times are obtained by
\begin{eqnarray}
    \mu_n &=& \tfrac{1}{n+1}[ \langle\tau_1\rangle +(n-1)\langle \tau_{i\neq 1,n+1}\rangle +\langle\tau_{n+1}\rangle],\\
    \sigma^2_n &=& \tfrac{1}{n+1}[ \langle\tau_1^2\rangle +(n-1)\langle \tau_{i\neq 1,n+1}^2\rangle+ \langle\tau_{n+1}^2\rangle ]-\mu_n^2.
\end{eqnarray}
Then we calculate the coefficient of variation of interevent times as following:
\begin{equation}
    \label{eq:rn_obc}
    r_n(x,y,z) = \sqrt{\tfrac{2[x-(n-1)y][n+2-x+(n-1)y]}{n+2} +(n+1)[(1-x-z)^2+z^2+(n-1)y^2]-1}.
\end{equation}
\end{widetext}
Here we have defined
\begin{equation}
    x\equiv \tfrac{\Delta}{T},\ y\equiv \tfrac{\tau_0}{T},\ z\equiv \tfrac{t_0}{T},
\end{equation}
satisfying the conditions that 
\begin{equation}
    (n-1)y\leq x\leq 1-y,\ y\leq z\leq 1-x-y,\ y\leq \tfrac{1}{n+1}.
\end{equation}
The burstiness parameter for open boundary condition is obtained as $B_n(x,y,z)=\frac{r_n(x,y,z)-1} {r_n(x,y,z)+1}$. 

We discuss the reference cases. The regular time series may correspond to the case of $r_n=0$. It implies that all interevent times, including $\tau_1$ and $\tau_{n+1}$, must be the same, i.e., $x=\frac{n-1}{n+1}$ and $y=z=\frac{1}{n+1}$. The random time series is obtained when $x=1$ and $y=z=0$, where $z=0$ is needed to avoid any memory effects. Finally, the bursty limit for maximizing $r_n$ implies the situation when all events occur at the same time, i.e., $x=y=0$. In order to get the maximum value of $r_n$, we choose $z=0$. In sum, one gets
\begin{eqnarray}
    B_n(\tfrac{n-1}{n+1}, \tfrac{1}{n+1}, \tfrac{1}{n+1})&=&-1,\nonumber\\
    B_n(1,0,0)&=&\tfrac {\sqrt{n}-\sqrt{n+2}} {\sqrt{n}+\sqrt{n+2}}, \\
    B_n(0,0,0)&=&\tfrac {\sqrt{n}-1}{\sqrt{n}+1}.\nonumber
\end{eqnarray}
These results can be also obtained from those for the periodic boundary condition by replacing $n$ by $n+1$, because we have one more interevent time under the open boundary condition. Then, the novel definition of the burstiness measure for the open boundary condition reads
\begin{equation}
    A_n(r)=\tfrac{\sqrt{n+2}r-\sqrt{n}}{(\sqrt{n+2}-2)r+\sqrt{n}}
\end{equation}
for $0\leq r\leq \sqrt{n}$.

Finally, we study the effect of minimum interevent time, $\tau_{\rm min}$ or $\tilde y \equiv \frac{\tau_{\rm min}}{T}$, on the burstiness parameter. The bursty limit case is obtained when $x=(n-1)\tilde y$ and $z=\tilde y$. The random case is obtained when $x=1-\tilde y$ and $z=\tilde y$. For these two cases, we use Eq.~(\ref{eq:rn_obc}). As for the regular case we consider a specific time series that one interevent time is $\tau_{\rm min}$, while all other $n$ interevent times are the same as $\frac{T-\tau_{\rm min}}{n}$. Here $z$ can be either $\tilde y$ or $\frac{1-\tilde y}{n}$, leading to the same result for the coefficient of variation of interevent times as $r^*_n(\tilde y)=\frac{1-(n+1)\tilde y}{\sqrt{n}}$. Then, the calculation of novel burstiness measure $A_{n,\tilde y}(r)$ for the open boundary condition is straightforward using the following conditions:
\begin{eqnarray}
    A_{n,\tilde y}[r^*_n(\tilde y)]&=&-1,\nonumber\\
    A_{n,\tilde y}[r_n(1-\tilde y,\tilde y,\tilde y)]&=&0,\\
    A_{n,\tilde y}[r_n((n-1)\tilde y,\tilde y,\tilde y)]&=&1.\nonumber
\end{eqnarray}

\section{General formula of the novel burstiness measure}\label{sect:generalA}

We derive a general formula of the novel burstiness measure $A(r)=\frac{ar-b}{r+c}$ with coefficients $a$, $b$, and $c$ when the conditions for reference cases are given as following:
\begin{eqnarray}
    A(r_-)&=&-1,\nonumber\\
    A(r_0)&=&0,\\
    A(r_+)&=&1.\nonumber
\end{eqnarray}
Here $r_-$, $r_0$, and $r_+$ denote the coefficients of variation for reference cases, respectively, e.g., see Eq.~(\ref{eq:conditions_An}) or Eq.~(\ref{eq:conditions_Any}). Using these conditions, one gets
\begin{equation}
    \label{eq:generalA}
    A(r)=\tfrac{(r_+ - r_-)(r-r_0)}{(r_+ + r_- - 2r_0)r+(r_+ + r_-)r_0-2r_+r_-}.
\end{equation}

\end{document}